\documentclass[]{jpconf}
\usepackage{graphicx}
\usepackage{feynmp-auto}
\usepackage{amsmath}
\usepackage{float}
\usepackage{hyperref}
\usepackage{subfig}
\usepackage{xspace}
\usepackage{multirow}

\renewcommand{\pt}{\ensuremath{p_\text{T}}\xspace}

\begin{document}
\title{The Madala hypothesis with Run 1 and 2 data at the LHC}
\author{Stefan von Buddenbrock,$^a$ Alan S. Cornell,$^b$ Mukesh Kumar$^b$ and Bruce Mellado$^a$}
\address{$^a$School of Physics, University of the Witwatersrand, Wits 2050, South Africa}
\address{$^b$National Institute for Theoretical Physics; School of Physics and Mandelstam Institute for Theoretical Physics, University of the Witwatersrand, Johannesburg, Wits 2050, South Africa}

\ead{stef.von.b@cern.ch}

\begin{abstract}
The Madala hypothesis postulates a new heavy scalar, $H$, which explains several independent anomalous features seen in ATLAS and CMS data simultaneously. It has already been discussed and constrained in the literature by Run 1 results, and its underlying theory has been explored under the interpretation of a two Higgs doublet model coupled with a scalar singlet, $S$. When applying the hypothesis to Run 2 results, it can be shown that the constraints from the data are compatible with those obtained using Run 1 results.
\end{abstract}

\section{The Madala hypothesis\label{sec:intro}}

Searches for physics beyond the Standard Model (BSM) have become ubiquitous since the discovery of the Standard Model (SM) Higgs boson, $h$. The discovery of the Higgs boson was the first step in understanding the nature of electroweak symmetry breaking (EWSB). There are currently a plethora of BSM physics scenarios extending the notion of EWSB in the literature, many of these being within the reach of the Large Hadron Collider (LHC) running at an energy of $\sqrt{s}=13$~TeV.

\begin{figure}
\centering
\subfloat[]{\includegraphics[width=0.3\textwidth]{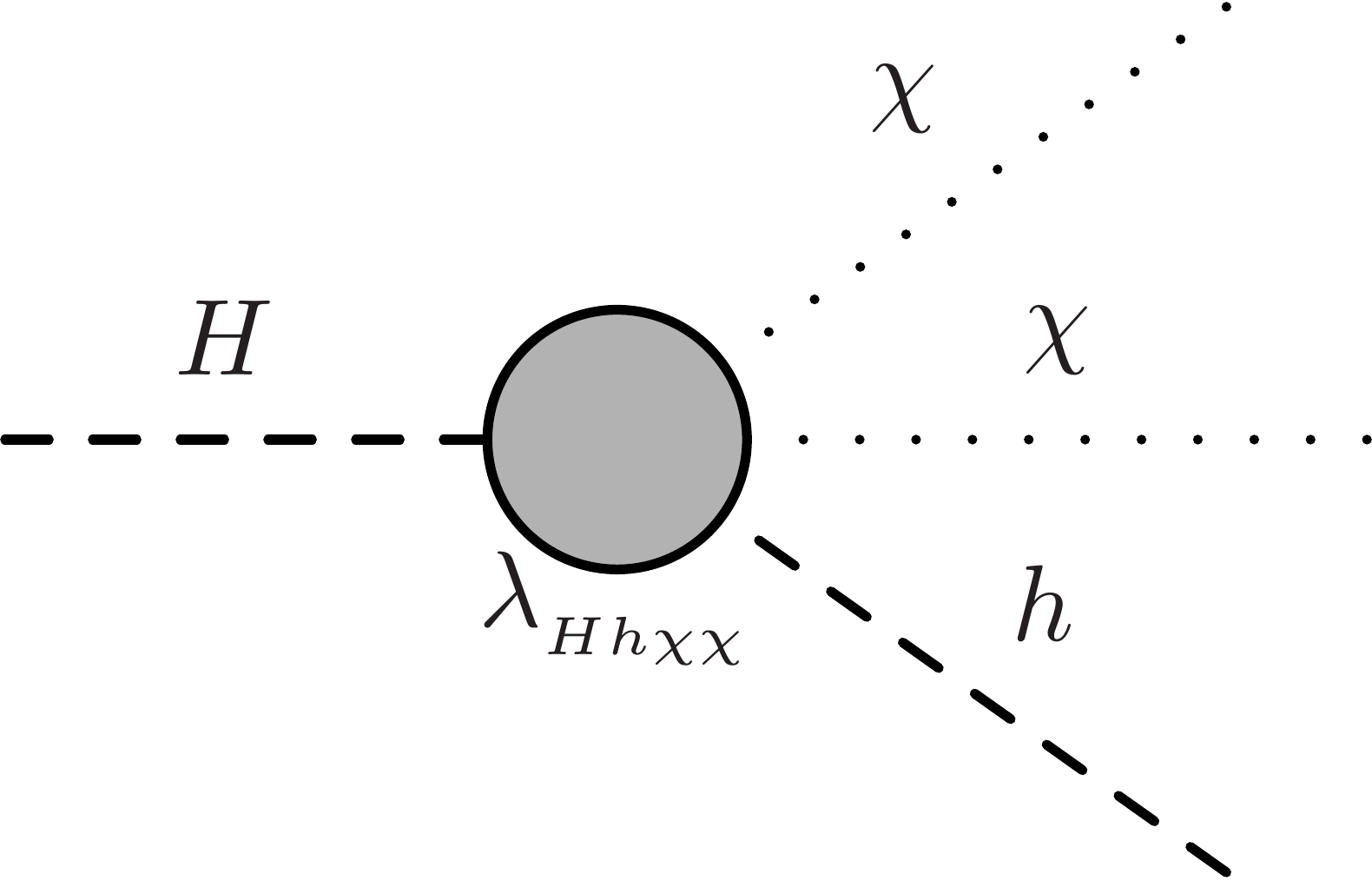}}
\quad\quad\quad\quad\quad
\subfloat[]{\includegraphics[width=0.3\textwidth]{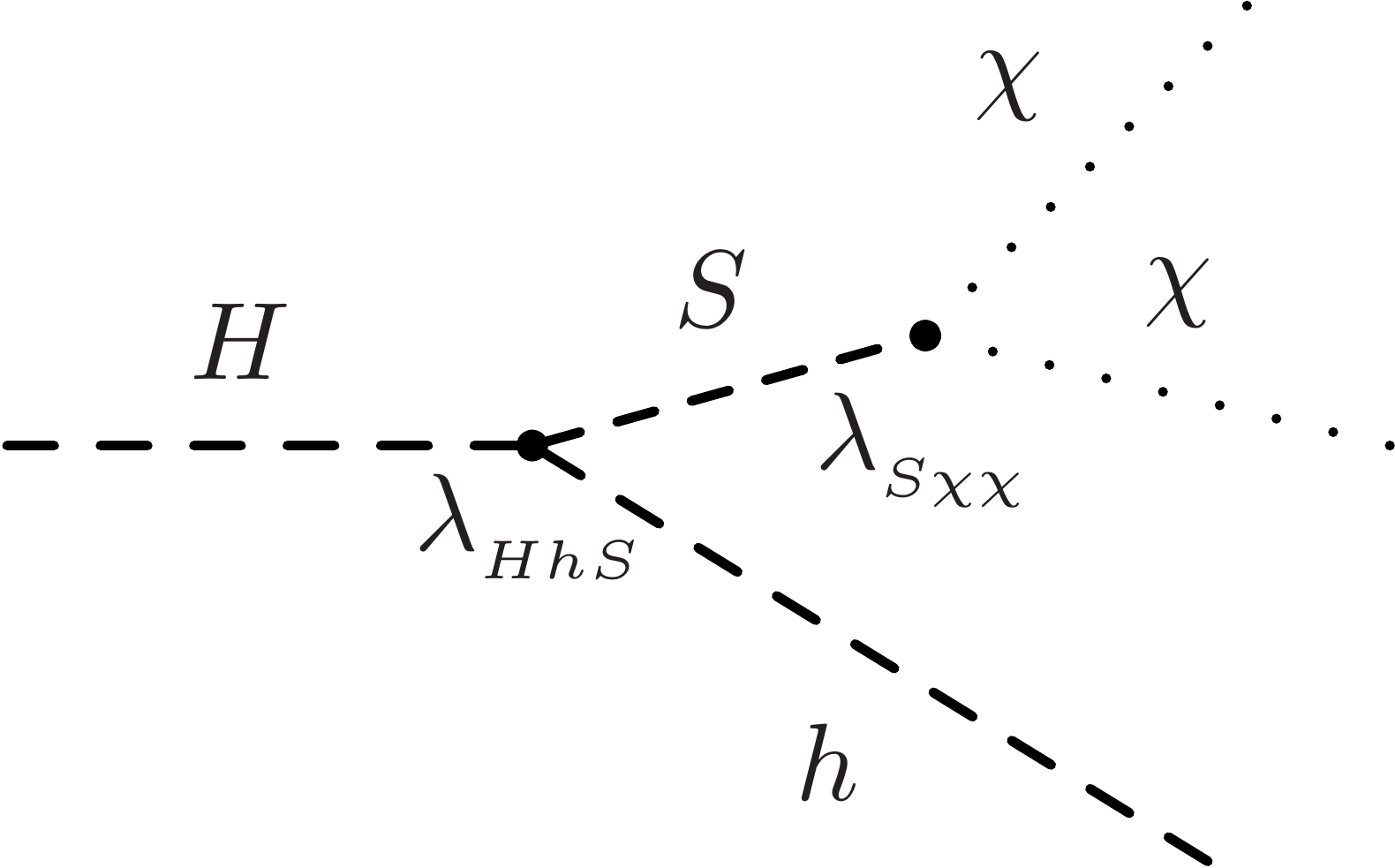}}
\caption{Higgs production in association with DM through the decay of $H$ using (a) an effective vertex and (b) the decay of a DM mediator $S$.}
\label{fig:diagrams}
\end{figure}

One such scenario is the \textit{Madala hypothesis}. The Madala hypothesis was formulated in 2015 with the aim of connecting and explaining several anomalies in the data from Run 1 of the LHC~\cite{vonBuddenbrock:2015ema}. At first, this was done through the introduction of a heavy boson $H$ -- the \textit{Madala boson} -- with a mass in the range $2m_h<m_H<2m_t$. It was shown that if the SM Higgs boson could be produced via the decay of $H$, it would contribute to the apparent distortion of the Higgs \pt spectrum from the Run 1 ATLAS differential distributions~\cite{Aad:2014lwa,Aad:2014tca}. This is achieved through the effective decay vertex of $H\to h\chi\chi$ shown in \autoref{fig:diagrams}(a), where $\chi$ is a scalar dark matter (DM) candidate of mass $\sim \tfrac{1}{2}m_h$. A fit was performed using a complete set of ATLAS and CMS data (at the time), and the parameters of the model were constrained. The two parameters of interest which were are constrained were the mass of $H$, to a value of $m_H=272^{+12}_{-9}$~GeV, and the scaling factor $\beta_g=1.5\pm0.6$ which modifies the gluon fusion ($gg$F) production cross section of $H$.

The hypothesis was then extended by introducing a DM mediator $S$~\cite{vonBuddenbrock:2016rmr}, such that the effective decay vertex in \autoref{fig:diagrams}(a) is resolved into the cascading decay process depicted in \autoref{fig:diagrams}(b). The mass of $S$ lies in the range $m_h<m_S<m_H-m_h$ in order for the decay process $H\to Sh$ to be on-shell.\footnote{The $S$ boson is not actually required to decay on-shell from $H$; this is merely a convenient mass range to simplify phenomenology. This facet has earned it the nickname \textit{the Shelly boson}.} The $H$ then preferentially decays to one of three pairs of bosons: $H\to SS,Sh,hh$. The $S$ boson was also considered to have small couplings to SM particles, and the assumption that was made is that it can be Higgs-like such that all of its branching ratios (BRs) are already defined. The $S$ would then decay predominantly to the vector bosons $Z$ and $W^\pm$ if it has a mass of around 160~GeV or above.

Since these studies were done, however, several new results have been released by the ATLAS and CMS collaborations. In particular, the first Run 2 results for resonant di-Higgs production and di-boson production have been released. In addition to this, both ATLAS and CMS have published Run 1 differential distributions (including Higgs \pt) for the $h\to WW\to e\nu\mu\nu$ channel~\cite{Aad:2016lvc,Khachatryan:2016vnn}, and ATLAS have also released their Run 2 result for $h\to\gamma\gamma$~\cite{ATLAS:2016nke}. It is therefore instructive to determine whether or not these new results are compatible with the Run 1 fit result mentioned above.

\section{Exploring resonant search channels\label{sec:searches}}

The first check for compatibility is to explore the resonant search channels which contributed to the Run 1 fit result mentioned above. These include resonant searches for $H\to hh$ and $VV$ (where $V=Z,W^\pm$).

Since no combination of these results exists, extracting meaningful information from them becomes a matter of statistics. In this study, results are combined through the addition of units of $\chi^2$. For measurements, $\chi^2$ is calculated as Pearson's test statistic. That is, a measurement $\mu^\text{exp}$ along with its associated uncertainty $\Delta\mu^\text{exp}$ are tested against a theoretical prediction $\mu^\text{th}$ and its uncertainty $\Delta\mu^\text{th}$ by calculating the following:\footnote{The denominator here differs from Pearson's test statistic, since it already assumes that the theoretical and experimental uncertainties are independent and can therefore be added in quadrature.}

\begin{equation}
\chi^2=\frac{(\mu^\text{th}-\mu^\text{exp})^2}{(\Delta\mu^\text{th})^2 + (\Delta\mu^\text{exp})^2}.
\label{eqn:chisquare_measurement}
\end{equation}

Most results in the literature, however, come in the form of 95\% CL limits. In this case, a modified version of Pearson's test statistic is used. Namely, for an observed and expected limit $L^\text{obs}$ and $L^\text{exp}$, respectively, a theoretical prediction $\mu^\text{th}$ can be tested using the following:

\begin{equation}
\chi^2=\frac{(L^\text{obs}-L^\text{exp}-\mu^\text{th})^2}{(L^\text{exp}/1.96)^2},
\label{eqn:chisquare_limit}
\end{equation}
where the factor of 1.96 in the denominator arises from the fact that a 95\% CL corresponds to 1.96 units of standard deviation.

\begin{figure}
\centering
\vspace{-70pt}

\subfloat[]{\includegraphics[page=3,width=0.45\textwidth]{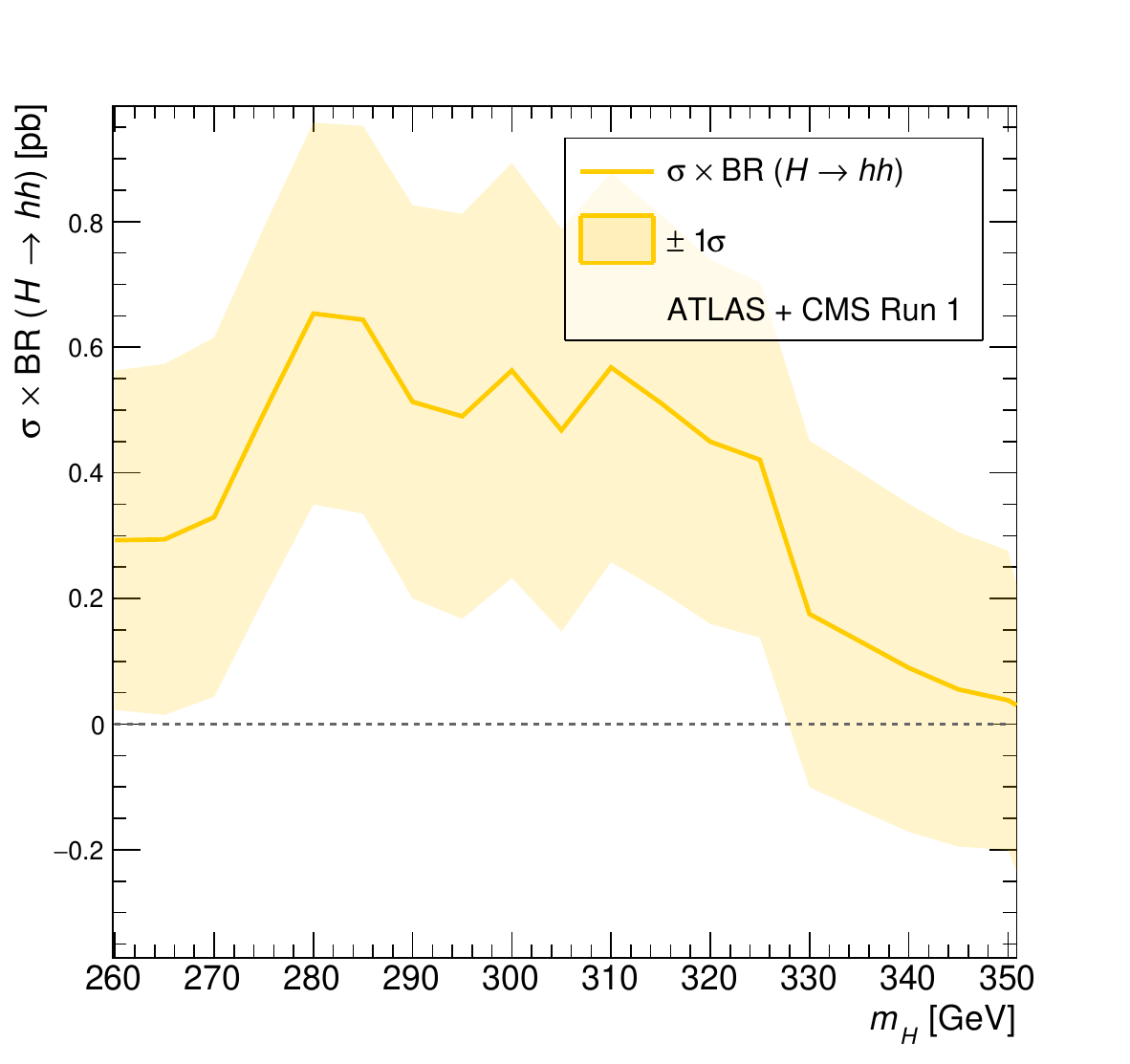}}
~
\subfloat[]{\includegraphics[page=6,width=0.45\textwidth]{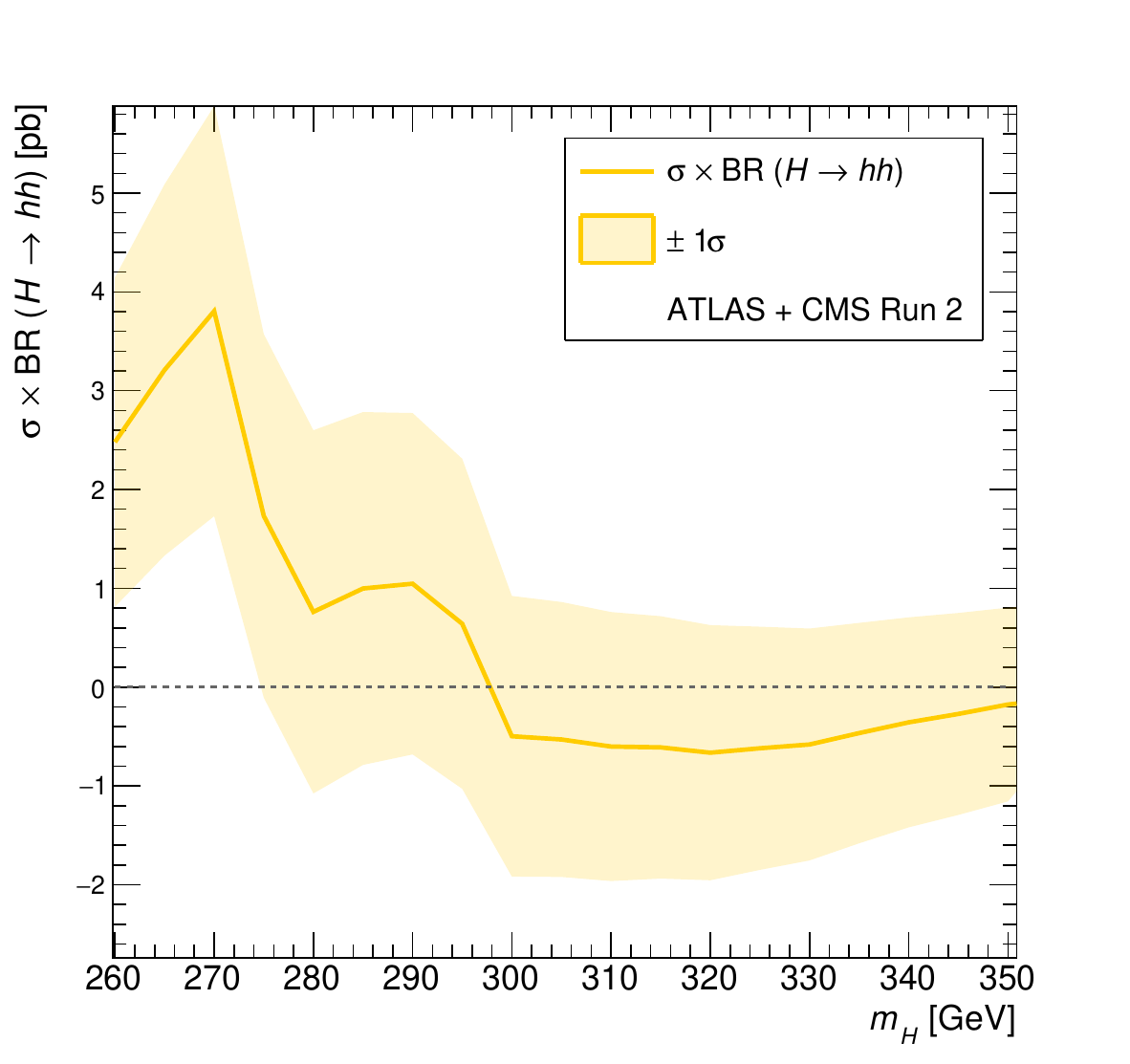}}
\vspace{-13pt}

\subfloat[]{\includegraphics[page=9,width=0.45\textwidth]{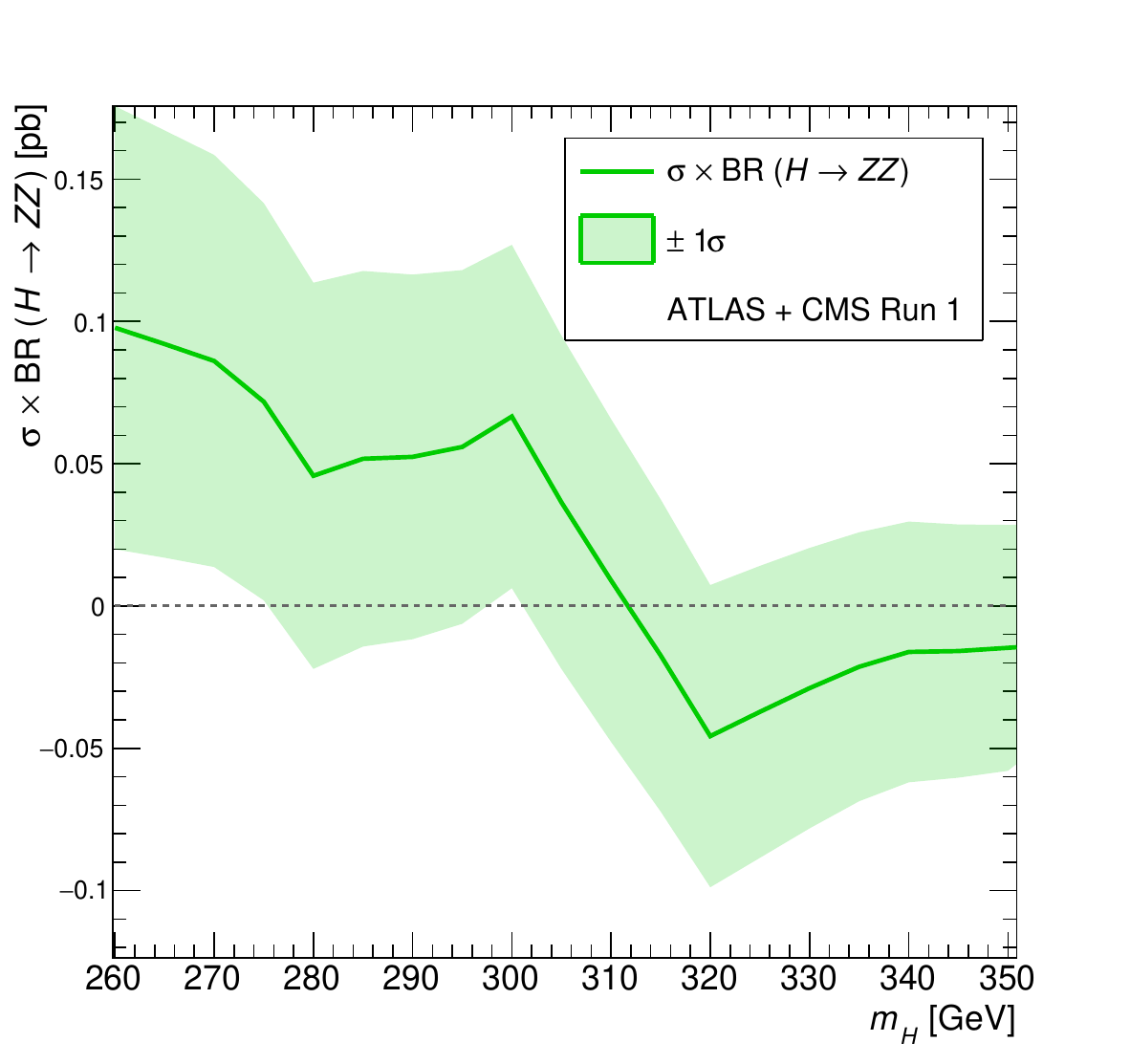}}
~
\subfloat[]{\includegraphics[page=12,width=0.45\textwidth]{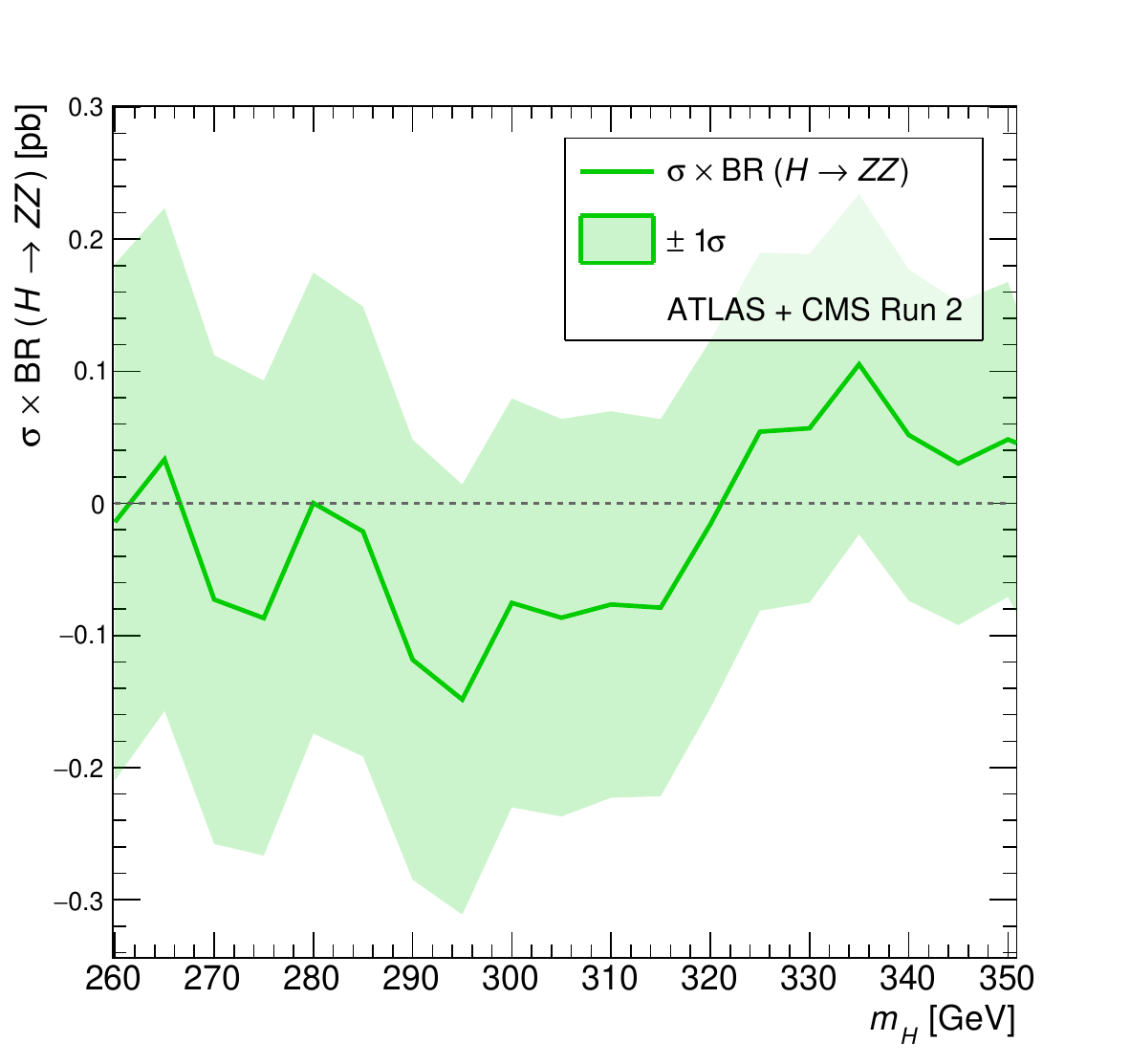}}
\vspace{-13pt}

\subfloat[]{\includegraphics[page=15,width=0.45\textwidth]{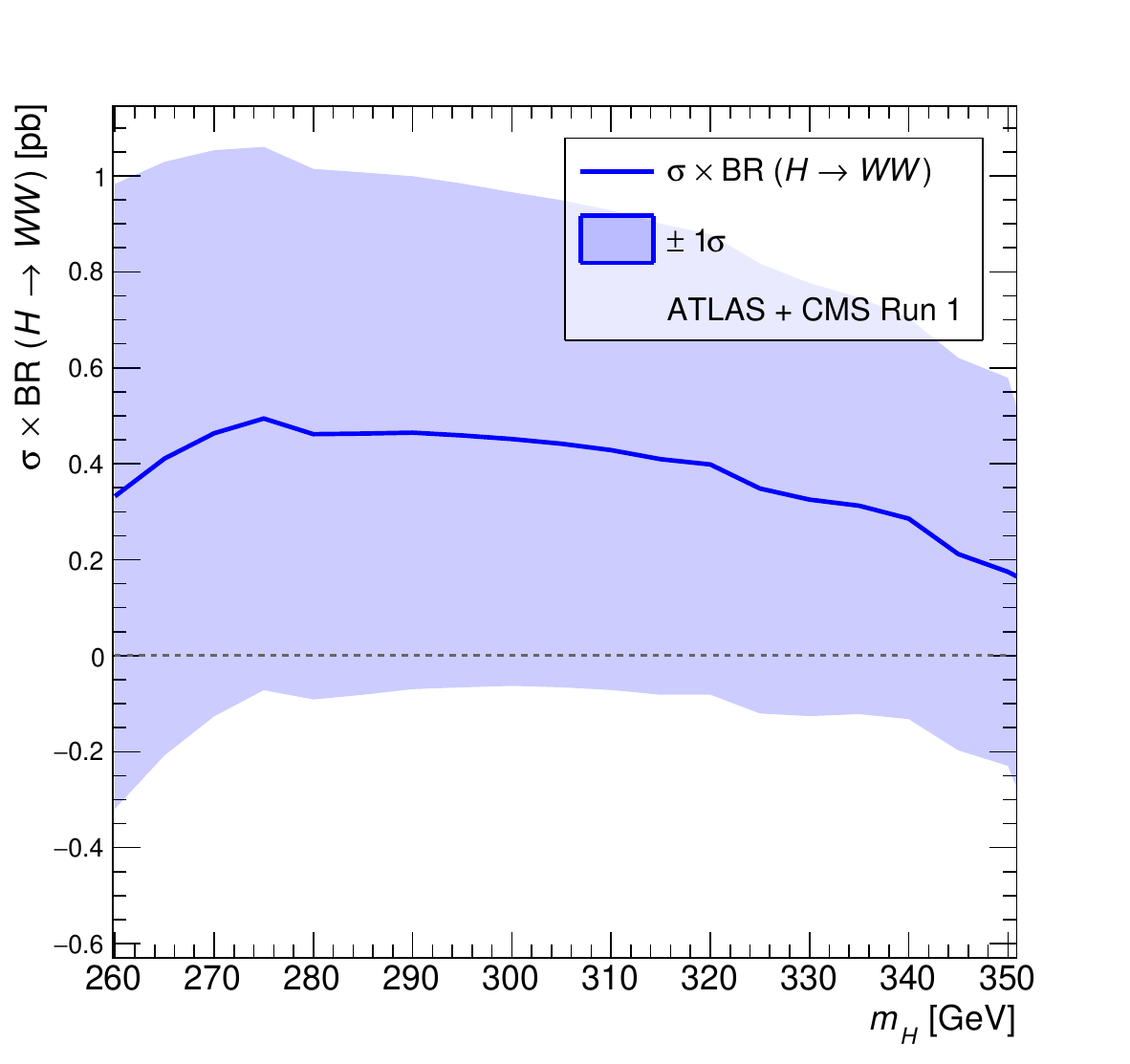}}
~
\subfloat[]{\includegraphics[page=18,width=0.45\textwidth]{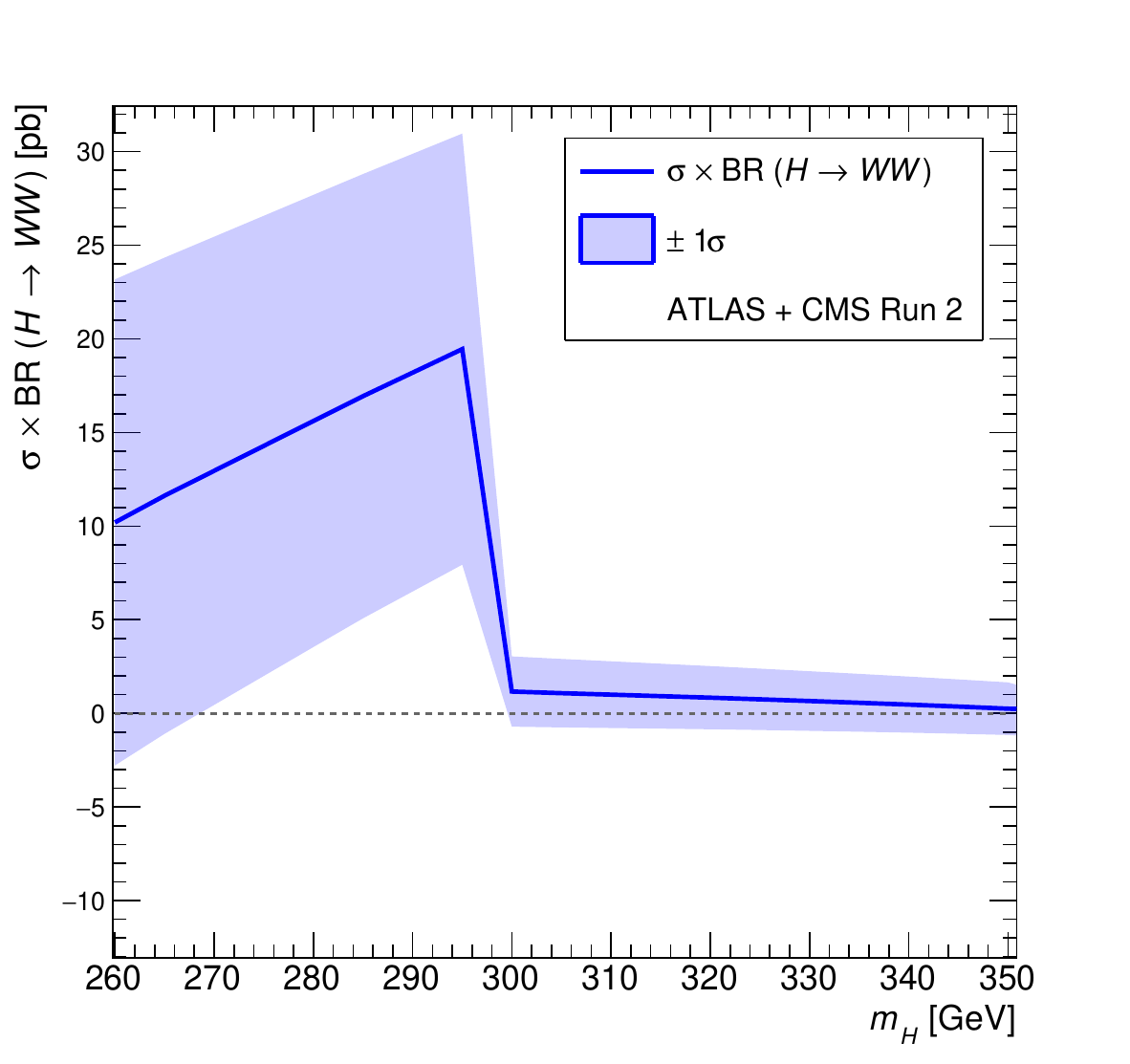}}
\caption{The best fit values of cross section times BR for $H$ production and decay into di-Higgs (top), $ZZ$ (middle), and $WW$ (bottom). The values have been separated into the 8~TeV Run 1 results (left) and the 13~TeV Run 2 results (right). The results on the left and right only include the data exclusively from Run 1 and Run 2, respectively, since the cross sections can not be directly compared at different energies.}
\label{fig:searches}
\end{figure}

To understand whether or not a potential signal already lies in the LHC data, the limits coming from the di-Higgs and di-boson searches listed in \autoref{tbl:searches} were scanned and evaluated using \autoref{eqn:chisquare_limit}. A best fit value for a production cross section times BR of $H$ was determined as a function of $m_H$ by minimising the sum of $\chi^2$ coming from each search channel. These best fit values as well as 1$\sigma$ error bands are shown in \autoref{fig:searches}. Note here that since results are shown both at 8 TeV and 13 TeV, the cross sections should not be directly compared between the two energies. However, the reader should keep in mind that the scaling factor from 8 TeV to 13 TeV for the production cross section of $H$ lies between 2.7 and 3.0 as $m_H$ increases from 250 GeV to 350 GeV (calculated using the NNLO+NNLL cross sections in reference~\cite{deFlorian:2016spz}). The plots indicate that the best fit value tends to deviate from the null hypothesis (i.e. that $H$ is not produced at all) mostly in the range between 260~GeV and 300~GeV. The only clear exception is in the Run 2 $H\to ZZ$ search results, where the best fit line underestimates even the null hypothesis. However, the $H\to ZZ$ BR was found to be small in the Run 1 fit result~\cite{vonBuddenbrock:2015ema}, and the value presented here is compatible with a small BR within the large uncertainties that surround the central value. This is compatible with the best fit mass of 272~GeV which was obtained in the Run 1 fit result mentioned in \autoref{sec:intro}. The Madala boson of course can be as light as 250 GeV, but since di-Higgs and di-boson results seldom consider masses below 260 GeV, the scans in \autoref{fig:searches} are limited.

\begin{table}
\vspace{-50pt}
\renewcommand{\arraystretch}{1.15}
\centering
\begin{tabular}{|c|c|c|c|c|}
	\hline
		\textbf{Result type} & \textbf{Collaboration} & \textbf{Run} & \textbf{Final state} & Luminosity [fb$^{-1}$] \\
		\hline
		Higgs $p_\text{T}$  & ATLAS & 1 & $\gamma\gamma$, $ZZ^*\to4\ell$, $WW^*\to e\nu\mu\nu$ & 20.3 \\
		\cline{2-5}
		spectrum  & CMS & 1 & $\gamma\gamma$, $ZZ^*\to4\ell$, $WW^*\to e\nu\mu\nu$ & 19.4-19.7 \\
		\cline{2-5}
		 & ATLAS & 2 & $\gamma\gamma$ & 13.3 \\
		 \hline
		 Di-Higgs & ATLAS & 1 & $bb\tau\tau$, $\gamma\gamma WW^*$, $\gamma\gamma bb$, $bbbb$ & 20.3 \\
		 \cline{2-5}
		 & CMS & 1 & $bb\tau\tau$, $\gamma\gamma bb$, multilepton & 19.5-19.7 \\
		 \cline{2-5}
		 & ATLAS & 2 & $\gamma\gamma bb$ & 3.2 \\
		 &  &  & $bbbb$, $\gamma\gamma WW$ & 13.3 \\
		 \cline{2-5}
		 & CMS & 2 & $bbbb$, $bbWW$ & 2.3 \\
		 &  &  & $\gamma\gamma bb$ & 2.7 \\
		 &  &  & $bb\tau\tau$ & 12.9 \\
		 \hline
		 Di-boson & ATLAS & 1 & $WW$, $ZZ$ & 20.3 \\
		 \cline{2-5}
		  & CMS & 1 & $WW$, $ZZ$ & 19.7 \\
		 \cline{2-5}
		  & ATLAS & 2 & $ZZ\to4\ell$ & 14.8 \\
		  &  &  & $ZZ\to2\ell2\nu$ & 13.3 \\
		  &  &  & $WW\to e\nu\mu\nu$ & 13.2 \\
		  \cline{2-5}
		  & CMS & 2 & $ZZ\to2\ell2\nu$, $WW\to 2\ell2\nu$ & 3.2 \\
		 \hline
	\end{tabular}
	\caption{A list of results that are relevant for consideration in the Madala hypothesis. Since results are constantly being updated in newer publications from the experimental collaborations, the integrated luminosity that has been used in this paper has also been tabulated.}
	\label{tbl:searches}
\end{table}

\section{Fitting the Higgs \pt spectrum\label{sec:higgspt}}

Another key aspect of the Run 1 fit result mentioned in \autoref{sec:intro} is the Madala hypothesis's ability to predict a distorted Higgs \pt spectrum. In the Run 1 data, this effect is most notably seen in the ATLAS results, where differential distributions are presented in fiducial volumes of phase space~\cite{Aad:2014lwa,Aad:2014tca,Aad:2016lvc}. Through the effective decay of $H$ to $h\chi\chi$, the BSM component of Higgs \pt can be added to a SM prediction to reproduce the systematic enhancement of fiducial cross section in the range 20~GeV $<\pt<$ 100~GeV, therefore improving the theoretical description of the data.

\begin{table}
\vspace{-50pt}
\renewcommand{\arraystretch}{1.23}
\centering
	\begin{tabular}{|c|c|c|c|}
	\hline
	$p_\text{T}$ spectrum & Data period & Luminosity [fb$^{-1}$] & $\beta_g$ \\
	\hline
	ATLAS $h\to WW$ & Run 1 & 20.3 & $1.4\pm0.6$ \\
	ATLAS $h\to\gamma\gamma$ & Run 2 & 13.3 & $1.0\pm0.9$ \\
	CMS $h\to WW$ & Run 1 & 19.4 & $0$ \\
	\hline
	\end{tabular}
	\caption{Fit results for the study done on the Higgs \pt spectrum. Here the effective vertex shown in \autoref{fig:diagrams}(a) was used, with $m_H=270$~GeV and $m_\chi=60$~GeV.}
	\label{tbl:results}
\end{table}

\begin{figure}[b]
\subfloat[]{\includegraphics[page=10,width=0.5\textwidth]{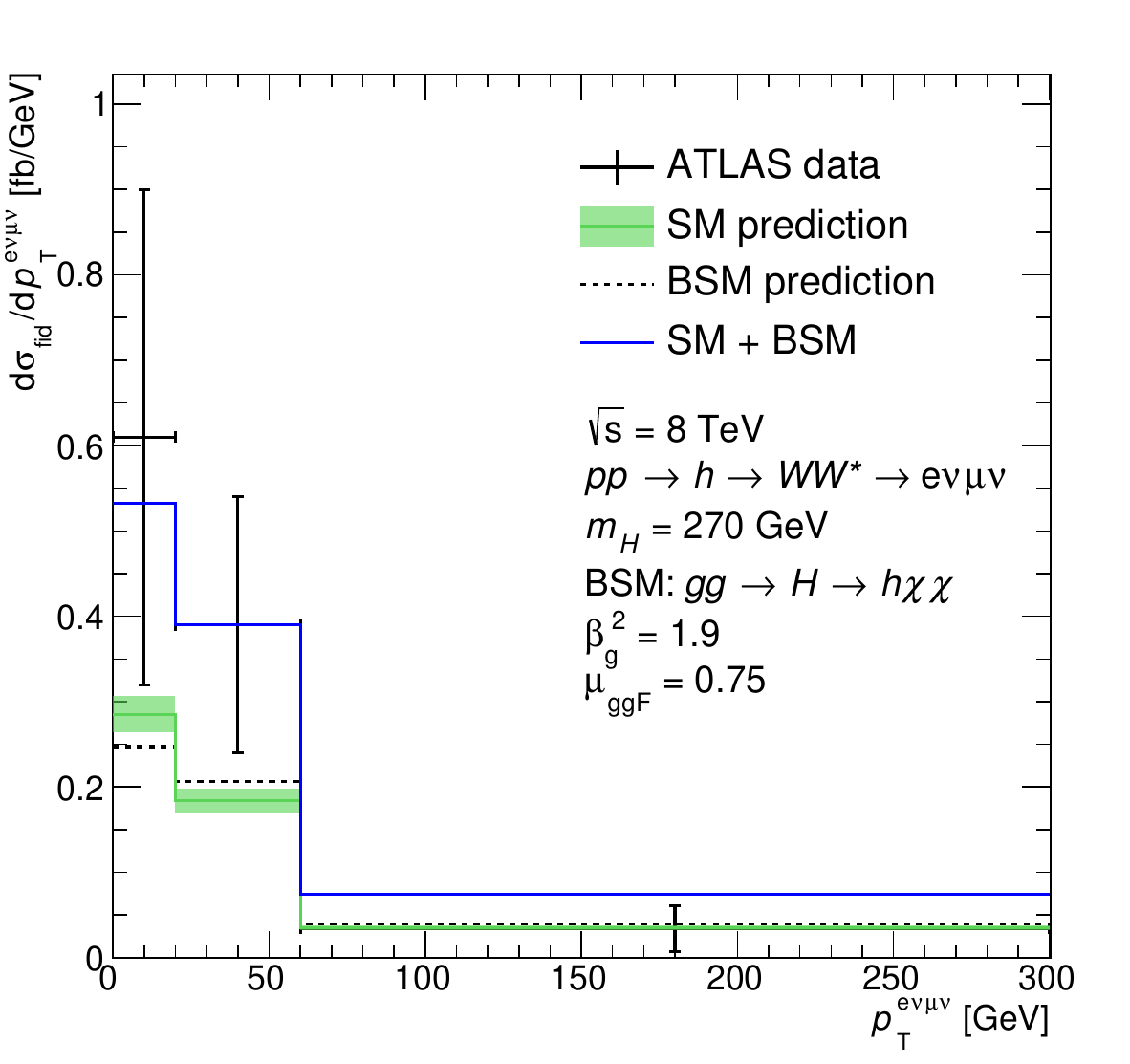}}
~
\subfloat[]{\includegraphics[page=11,width=0.5\textwidth]{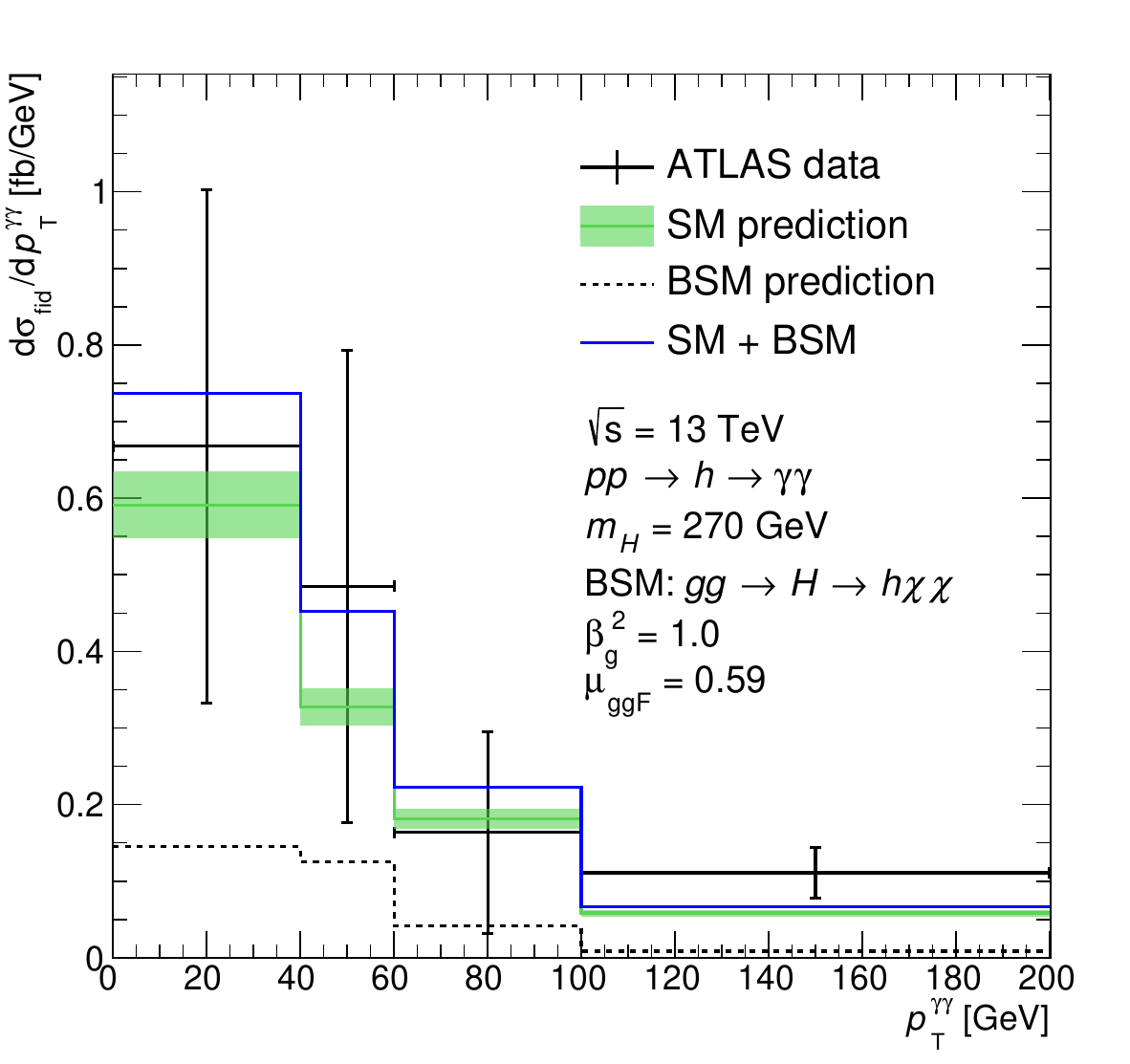}}
\caption{The Higgs \pt spectra for (a) the ATLAS Run 1 $h\to WW$ channel and (b) the ATLAS Run 2 $h\to\gamma\gamma$ channel. The mass points considered here are $m_H=270$~GeV and $m_\chi=60$~GeV, as described in the text.}
\label{fig:higgspt}
\end{figure}

In order to test whether or not such an improvement can be seen in the results released since the Run 1 fit result, a set of Monte Carlo (MC) samples were made to reproduce the different components of the Higgs \pt spectrum. The SM Higgs \pt spectrum was separated into its different production mechanisms. The $gg$F spectrum was generated using the NNLOPS procedure~\cite{Hamilton:2015nsa}, which is accurate to next-to-next-to leading order (NNLO) in QCD. The associated production modes -- vector boson fusion (VBF), $Vh$ and $tth$ which are commonly labelled together as $Xh$ -- were generated at next to leading order (NLO) using \textsc{MG5\_aMC@NLO}~\cite{Alwall:2014hca}. These spectra are scaled to the cross sections provided by the LHC Higgs Cross Section Working Group (LHCHXSWG)~\cite{deFlorian:2016spz} (from which the theoretical uncertainty also comes). The events are also passed through an even selection identical to the fiducial selection recommended by the experimental collaborations. A further scaling factor was applied to the SM $gg$F prediction, this being the reported signal strength of $gg$F (often denoted as $\mu_{gg\text{F}}$).

The BSM prediction (i.e. the Madala hypothesis prediction of $gg\to H\to h\chi\chi$ as shown in \autoref{fig:diagrams}(a)) was generated using \textsc{Pythia 8.2}~\cite{Sjostrand:2014zea}. These events were scaled to the LHCHXSWG N$^3$LO $gg$F cross sections for a high mass Higgs-like scalar, and passed through the fiducial selections as well. Since the Run 1 fit result had a best fit mass of $m_H=272$~GeV with $m_\chi=60$~GeV, the mass points considered for this study were $m_H=270$~GeV and $m_\chi=60$~GeV.

With the MC samples scaled accordingly each spectrum was added, and a $\chi^2$ value was calculated for each bin per channel, as in \autoref{eqn:chisquare_measurement}. The BSM component was scaled such that the total $\chi^2$ was minimised. This BSM scaling is interpreted to be equal to $\beta_g^2$, which is a dimensionless factor that multiplies the effective $g$-$g$-$H$ coupling, and therefore controls the production cross section of $H$ through $gg$F. The results of this fit are shown in \autoref{tbl:results}. The Run 1 fit result mentioned in \autoref{sec:intro} has a value of $\beta_g=1.5\pm0.6$, and here it can be seen that the ATLAS Run 1 $h\to WW$ and ATLAS Run 2 $h\to\gamma\gamma$ results are compatible with this value. The CMS Run 1 $h\to WW$ is not improved by the BSM hypothesis. The \pt spectra for the best fit values are shown in \autoref{fig:higgspt} for the two spectra which are improved by the BSM hypothesis.

\section{Conclusions\label{sec:conclusions}}


The Madala hypothesis was proposed in 2015 as an explanation of several anomalies in experimental data from the LHC. However, since then many newer results have come out which should corroborate the hypothesis if it exists in nature. In this work, these newer results have been tested using a statistical method, and are shown to be compatible with the result obtained in 2015. That is, most of the excesses from Run 1 which motivated the Madala hypothesis have reappeared in the current ensemble of Run 2 results.

However, this ensemble of Run 2 results comprises of preliminary studies which do not make use of the full integrated luminosity which has been accrued by the detectors over the duration of Run 2 at the LHC.
It is therefore imperative that a far more detailed study be done when such results become available, since the time is near when enough data will be available to make more definite statements about the Madala hypothesis. Some of the results in this paper are made from experimental plots containing even less than 5 fb$^{-1}$ of data. One would expect to be able to make a statement with $\mathcal{O}~\sim3\sigma$ confidence for individual search channels with at least 50 fb$^{-1}$ of data. Until such a time arrives, the phenomenology Madala hypothesis shall continue to be studied in the context of various BSM scenarios, to gain an understanding of how we might treat it in future.

\section*{References}

\bibliographystyle{iopart-num}
\bibliography{ref}
\end{document}